\documentclass[checkin,showpacs,aps,prl]{revtex4-1}

\usepackage{amsmath}
\usepackage{amssymb}
\usepackage{amsthm}

\newcommand{\br}{{\bf r}}

\newcommand{\bn}{\begin{enumerate}}
\newcommand{\en}{\end{enumerate}}
\newcommand{\ba}{\begin{eqnarray}}
\newcommand{\ea}{\end{eqnarray}}

\usepackage{graphicx,color}

\newcommand{\hp}{hyperpolarizability~}

\newcommand{\be}{\begin{equation}}
\newcommand{\ee}{\end{equation}}

\newcommand{\et}{{\it et al. }}
\newcommand{\ete}{{\it et al.}}
\newcommand{\torok}{T\"or\"ok }
\newcommand{\toroke}{T\"or\"ok}

\def\prl{{ Phys. Rev. Lett. }}

\renewcommand{\hp}{{\bf p}}

\newcommand{\Br}{{\bf r}}
\newcommand{\rt}{({\bf r},t)}

\begin{document}

\newcommand{\clr}{\color{red}}


\title{Electron transport under an ultrafast laser pulse: \\
  Implication
  for spin transport}

\author{Robert Meadows$^a$, Y. Xue$^b$, Nicholas Allbritton$^a$, 
and G. P. Zhang$^{a*}$}

 \affiliation{$^a$Department of Physics, Indiana State University,
   Terre Haute, IN 47809, USA}

 \affiliation{$^b$Milwaukee Area Technical College, 
Milwaukee, WI 53233-1443}

\date{\today}

\begin{abstract}
  {Laser-driven electron transport across a sample has garnered
    enormous attentions over several decades, because it potentially
    allows one to control spin transports in spintronics. But light is
    a transverse electromagnetic wave, how an electron acquires a
    longitudinal velocity has been very puzzling.  In this paper, we
    show a general mechanism is working. It is the magnetic field {\bf
      B} that steers the electron moving along the light propagation
    direction, while its strong transverse motion leads to local
    excitation.  We employ the formalism put forth by Varga and
    \toroke\ to show that if we only include {\bf E}, the electron
    only moves transversely with a large velocity.  Including both
    {\bf B} and {\bf E} and using real experimental laser parameters,
    we are able to demonstrate that a laser pulse can drive the
    electron along the axial direction by 20 to 262 $\rm \AA$,
    consistent with the experiments.  The key insight is that {\bf B}
    changes the direction of the electron and allows the electron to
    move along the Poynting vector of light. Our finding has an
    important consequence.  Because a nonzero {\bf B} means a
    spatially dependent vector potential ${\bf A} (\br,t)$, ${\bf
      B}=\nabla \times {\bf A}(\br,t)$, this points out that the
    Coulomb gauge, that is, replacing ${\bf A}(\br,t)$ by a spatial
    independent ${\bf A}(t)$, is unable to describe electron and spin
    transport under laser excitation.  Our finding is expected to have
    a potential impact on the ongoing investigation of laser-driven
    spin transport. }
\end{abstract}

\maketitle


Laser-induced electron and spin transport has attracted a broad
attention over several decades
\cite{brorson1987,boyen1997,hohlfeld2000,choi2014,vodungbo2012,pfau2012,kimling2017,hofherr2017,seifert2017,seifert2018,chen2019}.
In 1987, using an ultrashort laser pulse, Brorson \et
\cite{brorson1987} showed that the heat transport is very rapid across
the Au film, with the velocity close to the Fermi
velocity. Aeschlimann \et \cite{aeschlimann1996} estimated that the
excited electron velocity is $2\times 10^8$ cm/s and the inelastic
mean free path of 400 $\rm\AA$ in Au.  In 2011, Melnikov \et
\cite{melnikov2011} detected laser-excited spin-polarized carriers in
Au/Fe/MgO(001).
Transient spin current from ultrafast demagnetization in a
ferromagnetic Co/Pt layer can create a measurable spin accumulation in
Cu, Ag and Au \cite{choi2014b}.
Huisman \et \cite{huisman2017}, using THz emission spectroscopy,
detected spin currents in GdFeCo.  The temporal profile of spin pulses
indicates ballistic transport of hot electrons across the spin valves
\cite{alekhin2017}. Thermovoltages and tunneling magneto-Seebeck ratio
sensitively depend on the laser spot diameter \cite{martens2017}. Even
the exchange bias can be changed \cite{vallobra2017}. Hot electrons
alone can demagnetize a sample \cite{eschenlohr2013}.  Theoretically,
Lugovskoy and Bray \cite{lugovskoy1999} employed a Volkov wavefunction
to investigate electron dynamics.
Several studies adopted a
time-dependent temperature in the Fermi distribution
\cite{carpene2006,kratzer2022}, while others were based on the spin
superdiffusion transport \cite{battiato2010,rudolf2012,gupta2022}, but this is proved to be
  controversial \cite{lalieu2017,stamm2020}. Moisan \et
  \cite{moisan2014} found that hot electron spin dependent transfer
  between neighboring domains does not alter ultrafast
  demagnetization.  Including a laser field is possible in the
  Boltzmann equation, where nonequilibrium electron distribution can
  be treated \cite{bejan1997,brouwer2014,beens2022}.  In the atomistic
  spin simulation \cite{evans2014}, Vahaplar \et \cite{vahaplar2012}
  introduced a phenomenological optomagnetic field.
Fognini \et \cite{fognini2017} adopted a
  thermodynamic approach to ultrafast spin current, where laser
  pumping effect is encoded in the electron population change.
  Shokeen \et \cite{shokeen2017} contrasted spin flips and spin
  transport processes and found that spin flip dominates in Ni but in
  Co both play a role. This result is consistent with the experiment
  in Ni, where Schellekens \et \cite{schellekens2013} were unable to
  detect any major differences in Ni thin films on conducting and
  insulating substrates. An excellent review on hot-electron transport
  and demagnetization was given by Malinowski \et
  \cite{malinowski2018}.
So the central question
\cite{liu2023,kang2023} is why and how the electron could gain the
longitudinal velocity.

The goal of this paper is to investigate how a single laser pulse
could drive the electron in a thin film down the axial direction so
fast and so long.  We employ Varga and T\"or\"ok's method, which is
based on the Hertz vector and satisfies the Maxwell equation, to carry
out a series of simulations. We find a general conclusion. While the
electron oscillates strongly along the transversal direction and moves
very little, its axial velocity does not oscillate and is
unidirectional along the light propagation direction. Using an IR
pulse of 1.6 eV moves the electron along the axial direction by $z=20
\rm \AA$.  $z$ quadratically depends on the laser field amplitude and
linearly depends on the pulse duration. When we reduce the photon
energy $\hbar \omega$ to 0.1 eV, $z$ can reach over 120 $\rm \AA$,
Reducing $\hbar\omega$ to 0.06 eV, $z$ reaches 262 $\rm \AA$ even with
a weak pulse of $0.05 \rm V/\AA$, which is now closer to the
experimental observations \cite{brorson1987,melnikov2011}. Assuming
two different masses and two spin wavepacket spatial widths for
majority and minority spins, our model can describe the spin moment
change observed in experiments \cite{turgut2013}, without assuming
spin superdiffusion.  Central to our success is that although {\bf B}
is much weaker than {\bf E}, {\bf B} is the only one that can change
the direction of electron motion, a result that might be not
surprising to those in axial optics and plasma physics, but has been
unknown in laser-induced electron transport community. This welcoming
result requires a nonzero {\bf B}, but a nonzero {\bf B} subsequently
requires a spatially dependent vector potential ${\bf A}(\br,t)$,
owing to ${\bf B}=\nabla \times {\bf A}(\br,t)$. This has an important
ramification. The dipole approximation is frequently employed in model
and first-principles theories
\cite{jap11,krieger2015,chen2019w,prl20}, but in order to describe
electron transports in thin films, one has to use ${\bf A}(\br,t)$,
not ${\bf A}(t)$.  Our finding will have an important impact on future
ultrafast electron and spin transport
\cite{brorson1987,melnikov2011,stanciu2007,mplb18,aip20}.

\begin{figure}
    \centering
  \includegraphics[angle=0,width=0.5\columnwidth]{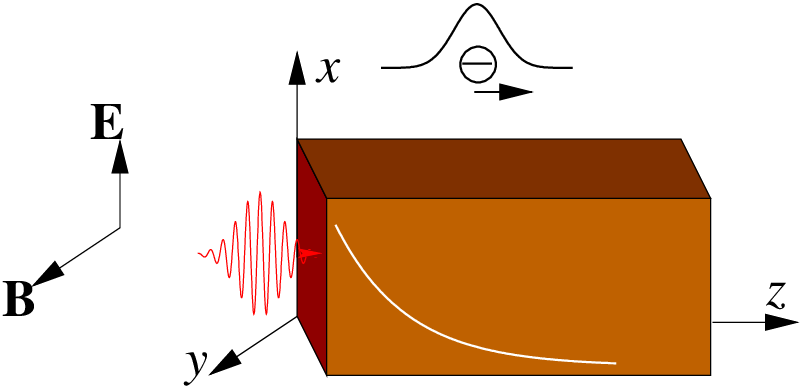}
  \caption{A linearly polarized laser pulse propagating along the $z$
    axis with the electric field ${\bf E}$ along the $x$ axis and the
    magnetic field {\bf B} along the $y$ axis. Inside the sample, both
    electric and magnetic fields reduce exponentially.
    The electron moves as a wavepacket along the $z$ axis. 
  }
    \label{fig1}
\end{figure}

A fundamental difference between the electric-field-induced transport
and laser-induced transport is that the latter does not have a voltage
bias along the carrier moving direction.  A light wave propagating
toward a sample along the $z$ axis has its electric-field {\bf E}
along the $x$ axis and the magnetic field {\bf B} along the $y$ axis.
Figure \ref{fig1} schematically illustrates a typical experimental
geometry. 

The Hamiltonian of the electron inside an electromagnetic field is \be
H=\frac{1}{2m_e}({\bf p}-q{\bf A}(\br,t))^2 +q\phi, \ee where ${\bf
  A}(\br,t)$ is the vector potential, $\phi$ is the scalar potential,
$q$ is the charge of the electron, and ${\bf p}$ is the electron
momentum.  The equation of motion is given in terms of the generalized
Lorentz force as \cite{griffiths2018}, \be m\frac{d^2\Br}{dt^2}=q {\bf
  E}\rt+\frac{q}{2m}(\hp\times {\bf B}\rt - {\bf B}\rt \times \hp)-
\frac{q^2}{m} ({\bf A}\rt\times {\bf B}\rt). \ee If we set ${\bf
  B}\rt$ to zero, then every term, except the first term, on the right
side, is zero, so the electron only moves along ${\bf E}\rt$. This
demonstrates that it is indispensable to include ${\bf B}\rt$.  In the
limit that {\bf B}, not ${\bf A}\rt$, is independent of space, one recovers
the usual Lorentz force, \be m\frac{d^2\Br}{dt^2}=q {\bf E}+q{\bf
  v}\times {\bf B}, \ee where the velocity  ${\bf v}$ is
defined through the mechanical momentum $ {\bf v}\equiv
\frac{{\bf p}-q{\bf A}(\Br,t)}{m}.$

We follow
Varga and \torok \cite{varga1998}, and start from the Hertz vector,
which satisfies the vectorial Helmholtz equation, so the resultant
electric and magnetic fields automatically satisfy the Maxwell
equation, 
given in an integral form as \be \textbf{E} = \frac{1}{2\pi k^2}
\int_{-\infty}^{\infty} \int_{-\infty}^{\infty} {\bf g}(k_x,
k_y)F(k_x,k_y)e^{i(k_xx+k_yy+k_zz)}dk_xdk_y, \ee and the magnetic
field is \be \textbf{B} = -\frac{i\sqrt{\epsilon\mu}}{2\pi
  k}\int_{-\infty}^{\infty}\int_{-\infty}^{\infty} \textbf{m}(k_x,
k_y)F(k_x, k_y)e^{i(k_xx+k_yy+k_zz)}dk_xdk_y, \ee where $ {\bf g}(k_x,
k_y) = (k^2-k_x^2)\hat{x}-k_xk_y\hat{y}-k_xk_z\hat{z},$ $
\textbf{m}(k_x, k_y) = k_z\hat{y}-k_y\hat{z},$ and $F$ is given by
\[ F(k_x, k_y) =
\frac{1}{2\pi}\int_{-\infty}^{\infty} \int_{-\infty}^{\infty}
V(x,y,0)e^{-i(k_xx+k_yy)}dxdy.\]
Here $V(x,y,0) =A_0
e^{-\frac{(x^2+y^2)}{w_0^2}}$  has the dimension of the
electric field, where $A_0$ is the field amplitude and $w_0$ is the pulse
spatial width chosen as $10\lambda$. $\lambda$ is the wavelength of
the pulse. 

Within 
the paraxial approximation, we 
integrate the above equations analytically to get the reduced
$\tilde{\bf E}$ and $\tilde{\bf B}$, \be
\begin{array}{lll}
\tilde{E}_x = \left[1+ \frac{4x^2-2w_0^2}{w_0^4k^2} \right],& 
\tilde{E}_y = \frac{4xy}{w_0^4k^2 },&  \tilde{E}_z =
-\frac{2ik_zx}{w_0^2k^2 }. 
\\ \tilde{B}_x=0,&
\tilde{B}_y= -\frac{ik_z\sqrt{\epsilon\mu}}{k},& \tilde{B}_z=
-\frac{2y\sqrt{\epsilon\mu}}{kw_0^2}. 
\end{array}
\label{varga1}
\ee
The final electric field and magnetic fields are \be {\bf E}= A_0
\tilde{\bf E}e^{ik_zz-i\omega t-\left
  (\frac{x^2+y^2}{w_0^2}\right)}, \hspace{0cm} {\bf B}= A_0 \tilde{\bf
  B}e^{ik_zz-i\omega t -\left
  (\frac{x^2+y^2}{w_0^2}\right)}. \label{varga3} \ee

When the light enters a sample, both fields are reduced by
$e^{-z/(2\lambda_{pen})}$, where $\lambda_{pen}$ is the penetration
depth and 2 comes from the fact that penetration depth is defined at
$1/e$ the incident fluence and the fluence is proportional to the
square of the electric field. $\lambda_{pen}$ is chosen to be 14 or 28
nm, typically values in fcc Ni.

 In our study, we choose a linearly $x$-polarized pulse that is
 propagating along the $z$ axis. The pulse is a Gaussian of duration
 $\tau$, amplitude $A_0$ and photon energy
 $\hbar\omega$. $A_0e^{-i\omega t}$ in ${\bf E}$ and ${\bf B}$ in
 Eq. \ref{varga3} is replaced by
 $A_0e^{-t^2/\tau^2}\cos(\omega t)$, so our electric and magnetic
 fields are 
 \ba{\bf E}\rt&=&
 A_0e^{-t^2/\tau^2}\cos(\omega t) \tilde{\bf
   E}e^{ik_zz-\frac{z}{2\lambda_{pen}}-\left
   (\frac{x^2+y^2}{w_0^2}\right)},\\ {\bf
   B}\rt&=&A_0e^{-t^2/\tau^2}\cos(\omega t) \tilde{\bf
   B}e^{ik_zz-\frac{z}{2\lambda_{pen}}-\left
   (\frac{x^2+y^2}{w_0^2}\right)}.
 \label{varga4}
 \ea These analytic forms of ${\bf E}$ and ${\bf B}$, which contain
 both the spatial and temporal dependences, greatly ease our
 calculation.  As a first step toward a complete transport theory, we
 treat the electron classically, and solve the Newtonian equation of
 motion numerically, \be \frac{d{\bf v}}{dt}=\frac{q}{m}\left [{\bf
     E}(\br,t)+{\bf v}\times {\bf B}(\br,t)\right ] -\frac{{\bf
     v}}{\Gamma},
\label{eq817b}
\ee where $m$ and $q$ are the electron mass and charge respectively,
${\bf v}$ is its velocity, and $\Gamma$ is a small damping to mimick
the resistance due to collision with other electrons and
nuclei. Although our method is classical, it fully embraces the real
space approach that is more suitable for transport, a key feature that
none of prior first-principles and model simulations is able to
achieve.  This will answer the most critical question whether the
electron can move along the axial direction.

\begin{figure}
\includegraphics[angle=0,width=1\columnwidth]{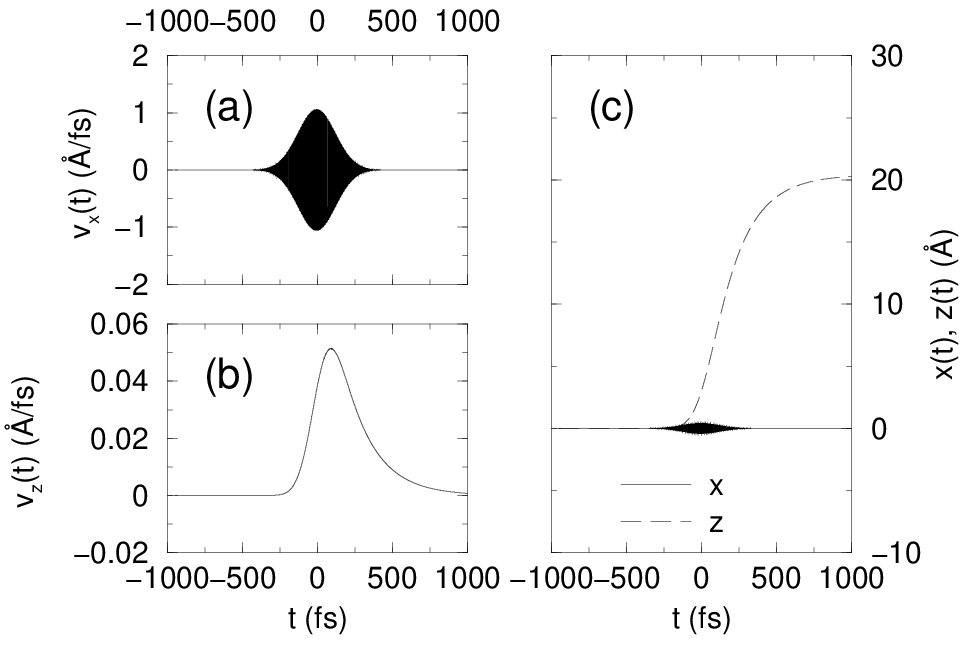}
\caption{ (a) The electron velocity $v_x$ strongly oscillates with
  time. Here we employ a pulse of $\hbar\omega=1.6$ eV, $\tau=180$ fs
  and $A_0=0.15\ \rm V/\AA$.  (b) $v_z$ does not have this
  oscillation. The peak around 100 fs is due to the damping
  $\Gamma$. (c) The electron moves little along the $x$ axis (solid
  line) and it oscillates locally. But along the $z$ axis, the
  electron clearly transports with a finite distance of about 20
  $\rm\AA$ (dashed line).  }
\label{fig2}
\end{figure}

We choose a pulse of $\hbar\omega=1.6$ eV, $\tau=180 $ fs,
$A_0=0.15\ \rm V/\AA$, often used in experiments \cite{jpcm10}. We set
the initial position and velocity of the electron to zero and  $\Gamma=
200\ \rm fs$.  Figure \ref{fig2}(a) shows that the velocity along the
$x$ axis, $v_x$, increases very rapidly, with the maximum velocity
reaching $\rm 1\AA/fs$, $10^{5}\rm m/s$, on the same order of
magnitude of the velocity found in Cu/Pt multilayers
\cite{bergeard2016}. Unfortunately, the experimental fluence was not
given, so a quantitative comparison is not possible. Nevertheless,
this velocity also agrees with another experiment in Co/Cu(001) films
\cite{wieczorek2015}.  Prior studies \cite{battiato2010} often use the
Fermi velocity to discuss spin transport, which is not
appropriate. Instead, one must obtain the actual electron velocity
first from the laser field, because in the absence of a laser field,
the sum of the velocities among the electrons must be zero and there
should be no transport. Although $v_x$ has a peak value, it oscillates
very strongly. By contrast, Fig.  \ref{fig2}(b) shows the velocity
$v_z$ increases with time, without oscillation. $v_z$ is always
positive, increases smoothly and peaks around 100 fs.  This peak time
is set by the laser parameter and the damping $\Gamma$.  Using a
larger $\Gamma$ leads to a larger $v_z$. The positivity of $v_z$,
regardless of the type of charge, is crucial to the electron
transport, and can be understood from the directions of ${\bf E}$ and
{\bf B}.  Suppose at one instant of time, {\bf E} is along the $+x$
axis and {\bf B} along the $+y$ axis. The electron experiences a
negative force along the $-x$ axis and gains the velocity along the
$-x$ axis, so the Lorentz force due to {\bf B} is along the $+z$
direction. Now suppose at another instance, {\bf E} changes to $-x$
and {\bf B} to $-y$, so the electron velocity is along $+x$, but the
Lorentz force is still along the $+z$ axis. If we have a positive
charge, the situation is similar. The fundamental reason why we always
have a positive force is because the light propagates along the $+z$
axis and the Poynting vector is always along $+z$ and ${\bf v}\times
{\bf B}$ points along the $+z$ axis. We test it with various laser
parameters and never find a negative $v_z$.  Under cw approximation,
Rothman and Boughn \cite{rothman2009a} gave a simple but approximate
expression for the dimensionless $v_z=\frac{1}{2}\left (
\frac{\omega_c}{\omega}\right)^2[\cos(\omega t)-1]^2$, and Hagenbuch
\cite{hagenbuch1977} gave $p_z=e^2A^2(\tau')/2mc$, both of which are
positive. Therefore, both their theories and our numerical results
with a pulse laser agree that the axial motion of the electron is
delivered by both {\bf E} and {\bf B}.  This is also consistent with
the radiation pressure from a laser beam can accelerate and trap
particles \cite{ashkin1970,ashkin1986}.

Figure \ref{fig2}(c) shows the displacement along the $x$ (solid line)
and $z$ (dashed line) axes, respectively. We see that $x$ does not
change very much. This shows that the electron excitation along the
$x$ axis is local. If we include the band structure of a solid, it
will stimulate both intraband and interband transitions \cite{jpcm23}
and demagnetize the sample locally. In our simple classical model,
these features cannot be included. But our study confirms the local
heating must occur. Besides the direct demagnetization due to heating,
it may stimulate magnon generation to destroy the magnetic long-range
ordering \cite{jap19}.  Central to our research is whether the
electron indeed moves along the axial direction.  Figure \ref{fig2}(c)
demonstrates clearly that the electron successfully moves along the
$z$ axis by 20 $\rm \AA$.

\begin{figure}
\includegraphics[angle=0,width=1\columnwidth]{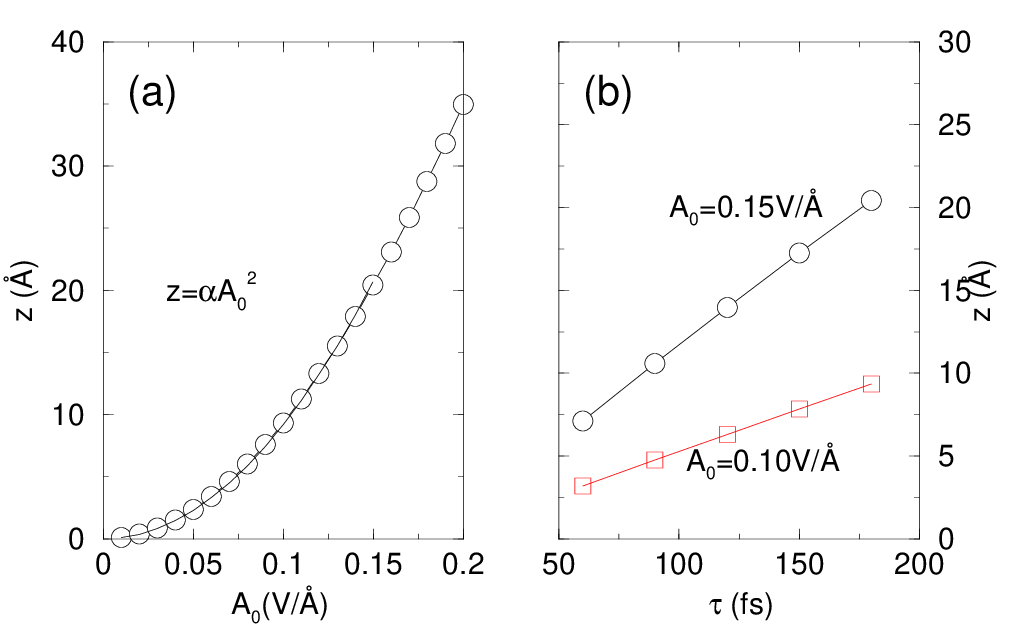}
\caption{ (a) The electron final position $z$ as a function of the
  laser amplitude $A_0$ (circles). The line is a fit to the data up to
  $A_0=0.15\rm V/\AA$, where $z=\alpha A_0^2$ and $\alpha=919.254 \rm
  \AA^3/V$.  Here $\hbar\omega=1.6$ eV and $\tau=180$ fs.  (b) $z$ as
  a function of $\tau$ for a fixed $\hbar\omega=1.6$ eV for
  $A_0=0.15\rm V/\AA$ (circles) and 0.10 $\rm V/\AA$ (squares). (c)
  $z$ as a function of $\hbar\omega$. 
}
\label{fig3}
\end{figure}

We can move one step further. Keeping the rest of laser parameters
unchanged, we change the laser field amplitude $A_0$ from 0.01 to 0.20
$\rm V/\AA$ and then compute $z$ for each amplitude. Figure
\ref{fig3}(a) is our result. First we notice that $z$ change is highly
nonlinear. We fit it to a quadratic function, $z=\alpha A_0^2$, up to
$A_0=0.15\rm V/\AA$, where $\alpha=919.254\rm \AA^3/V^2$, and find
that the fit is almost perfect. Because $|A_0|^2$ is directly
proportional to the fluence, this demonstrates $z$ is linearly
proportional to the laser fluence, which is exactly expected from the
Poynting vector ${\bf S}={\bf E}\times {\bf B}/c$. Thus, both
qualitatively and quantitatively our results can be understood. What
is less known is the dependence of $z$ on laser pulse duration
$\tau$. We change $\tau$ from 60 to 180 fs.  Figure \ref{fig3}(b)
shows $z$ increases with $\tau$ linearly. As expected, the increase is
steeper at $0.15 \rm V/\AA$ (the circles) than that
$A_0=0.10 \rm
V/\AA$ (boxes). The duration dependence leads us to wonder whether the
photon energy $\hbar\omega$ also affects the axial motion of the
electron.

We compute $z$ with our photon energy going from 0.02 up to 1.6 eV,
while keeping both the duration $\tau=180$ fs and amplitude
$A_0=0.05\rm V/\AA$ fixed.  Figure \ref{fig3}(c) shows an astonishing
result: $z$ is inversely proportional to $\hbar\omega$. At the lower
end of $\hbar\omega$, $z$ exceeds 100 $\rm \AA$. Note that at such a
low amplitude, a pulse of 1.6 eV only drives the electron by 2-3 $\rm
\AA$.  This explains why THz pulses become a new frontier for
ultrafast demagnetization \cite{shalaby2015,hudl2019,lee2021}. Polley \et
\cite{polley2018} employed a THz pulse to demagnetize CoPt films with
a goal toward ultrafast ballistic switching. Shalaby \et
\cite{shalaby2018} showed that extreme THz fields with fluence above
100 mJ/cm$^2$ can induce a significant magnetization dynamics in Co,
with the magnetic field becoming more important. 

Our result uncovers an
important picture. When the pulse oscillates more slowly, the electron
gains more grounds. Of course, a DC current can move electrons even
further, but then it does not have enough field intensity. This result
can be tested experimentally.

\begin{figure}
\includegraphics[angle=0,width=1\columnwidth]{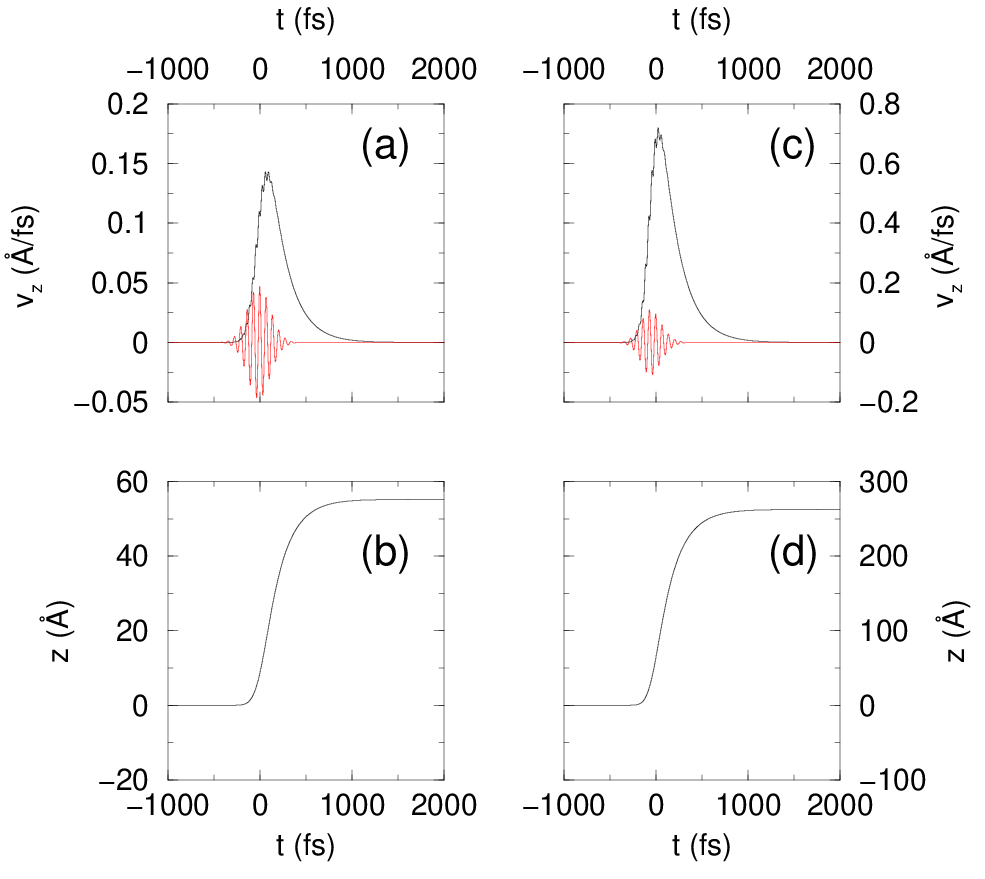}
\caption{ (a) $v_z$ as a function of time $t$. Our laser parameters
  are photon energy $\hbar\omega=0.06 $eV, duration $\tau=180$ fs,
  field amplitude $A_0=0.05 \rm V/\AA$. Inset: our laser pulse.  (b)
  $z$ as a function of $t$.  $z$ reaches 55.2 $\rm \AA$.  (c) $v_z$ as
  a function of time $t$ with $A_0=0.15 \rm V/\AA$. (d) The dependence
  of $z$ on $t$.  $z$ reaches 262 $\rm \AA$.  }
\label{fig4}
\end{figure}

We would like to see what happens to the electron under a THz pulse
excitation. We choose a pulse of $\hbar\omega=0.06$ eV, $\tau=180$ fs
and $A_0=0.05 \rm V/\AA$. Figure \ref{fig4}(a) shows that $v_z$
reaches 0.14 $\rm \AA/fs$, which is three times that with
$\hbar\omega=1.6$ eV and $A_0=\rm 0.15 V/\AA$ in
Fig. \ref{fig2}(b). Our laser pulse is shown in the inset of
Fig. \ref{fig4}(a). We should mention that our current $v_x$ has a
peak value of around 10 $\rm \AA/fs$, close to the Fermi velocity as
found in  the experiements \cite{brorson1987,melnikov2011}.
Figure \ref{fig4}(b) shows that $z$ reaches 55 $\rm\AA$.
Increasing $A_0$ to 0.15
$\rm V/\AA$ boosts $v_z$ by four times (Fig. \ref{fig4}(c)) and
increases  $z$ to 262 $\rm \AA$.  This distance is closer to the
experimental value. Melnikov \et \cite{melnikov2011} employed a
multilayer structure, (50 nm Au)/(15 nm Fe)/MgO(001). Once they
increase the thickness of Au to 100 nm, they find a delay signal in
second-harmonic generation. This shows the electron in Au can travel a
much longer distance \cite{razdolski2017}, suggesting that our
$\Gamma$ may be still too short. A much stronger effect is found in
plasma \cite{yang2011}.  This may indicate the free electron mass used
in our study may be too big.

\begin{figure}
\includegraphics[angle=0,width=0.5\columnwidth]{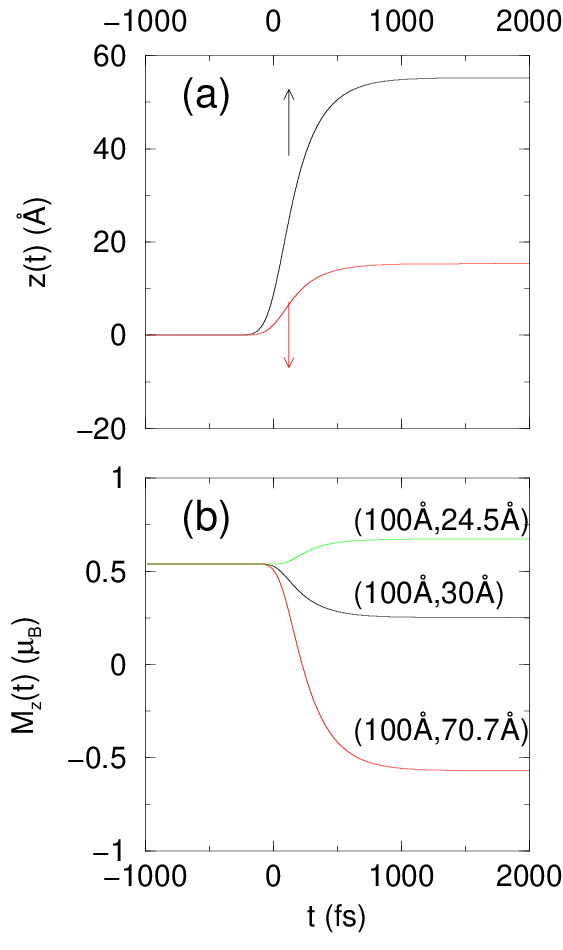}
\caption{ (a) $z$ for majority and minority spin channels as a
  function of time $t$, with $\hbar\omega=0.06 $ eV, duration
  $\tau=180$ fs, field amplitude $A_0=0.05 \rm V/\AA$. The majority
  and minority spin channels are mimicked by the effective mass of the
  electron. $m_{eff}^\uparrow=m_e$ and $m_{eff}^\downarrow=2m_e$.  (b)
  The spin moment changes as a function of time for three wavepackets
  whose widths are given on each curves.  }
\label{fig5}
\end{figure}

This opens the door to investigate the spin transport by using two
different effective masses of the electron.  Since the effective mass
of the majority spin is always smaller than that of the minority spin,
we assume $m_\uparrow=m_e$ and $m_\downarrow=2m_e$. Naturally, this
huge difference is somewhat exaggerated, but our intention is to see
whether this difference could produce something that can be related to
experiments.  With these different masses, we compute the $z_\uparrow$
and $z_\downarrow$ separately. We employ the same laser parameters as
above: $A_0=0.05 \rm V/\AA$, $\hbar\omega=0.06 $ eV, and $\tau=180$
fs.  For the same force, the acceleration $a_\uparrow$ is larger than
$a_{\downarrow}$, since $a=F/m$.  Figure \ref{fig5}(a) shows $z(t)$
for spin up and spin down. It is clear that $z_\downarrow$ is smaller
than $z_\uparrow$, as expected. This means that the majority spin
electron moves away from the origin earlier.

\newcommand{\zu}{z_{\uparrow}}
\newcommand{\zd}{z_{\downarrow}}

\newcommand{\wu}{w_{\uparrow}}
\newcommand{\Wd}{w_{\downarrow}}

The next question is how we can relate $z$ to the spin. We recall that
the classical position is always a good measure for the center of the
wavepacket of the electron in quantum mechanics.  In our mind, we
envision that a wavepacket of some size is centered around $z(t)$. As
time goes by, the center moves.  Henn \et \cite{henn2013} showed that
the excited electron spin spreads as a spin packet.  Since we do not
know the width of the wavepacket, we introduce them as a parameter. We
choose a simple one-dimensional Gaussian wavepacket, with the
wavefunction $\psi(z,t)=Ce^{-\beta z^2(t)}e^{ik_z z}$, where $\beta$
and $C$ are constants.  Other types should work as well.  We take fcc
Ni as our target example. The number of $3d$ electrons in the spin
majority channel is about 5, while that in the minority is 4.46. This
leads to the spin moment of 0.54 $\mu_B$ \cite{kittel}. The many-body
density $\rho=|\psi|^2$ is related to the electron number, and then we
have the spin moment at location $z=0$ as \be
M_z(\mu_B)(t)=5e^{-(z_\uparrow^2(t))/w_\uparrow^2}-4.46e^{-(z_\downarrow^2(t))/w_\downarrow^2},
\ee where both $z_\uparrow$ and $z_{\downarrow}$ are referenced to
$z=0$ and $w_{\uparrow(\downarrow)}$ is the width of spin majority
(minority) wavepacket. This equation has a nice feature. At $t=0$,
$M_z$ is 0.54 $\mu_B$.  As the electrons move away from $z=0$, the
local spin moment change sensitively depends on $w_\uparrow$ and
$w_\downarrow$. It turns out that we cannot arbitrarily choose them,
because they immediately lead to an unphysical spin moment change. Too
small values cannot reproduce experimental results.  A positive
$M_z>0$ means \be
e^{+(z_\uparrow^2(t))/w_\uparrow^2-(z_\downarrow^2(t))/w_\downarrow^2}>4.46/5
\rightarrow \frac{\zu^2}{\wu^2}-\frac{\zd^2}{\Wd^2}<0.1142 .\ee For
$\wu=100\rm \AA,\Wd=24.5\rm \AA$, Fig. \ref{fig5}(b) shows an increase
in $M_z$, which is very similar to Fig. 2(a) of
Ref. \cite{turgut2013}.  The reason is that when the majority
wavepacket has a larger width, the majority spin dominates over the
minority spin. Once we increase $\Wd$ to 30 $\rm \AA$, we see there is
a demagnetization, similar to Figs. 2(b), 2(c) and 2(d) of
Ref. \cite{turgut2013}, without assuming spin superdiffusion.
Increasing $\Wd$ further to $70.7 \rm \AA$ flips the spin to the
opposite direction, which resembles to all-optical spin switching
\cite{stanciu2007}. Since the electron scattering and other
scatterings are taken into account through $\Gamma$, our results are
applicable to solids.  Our finding reminds us the earlier study by von
Korff Schmising \et \cite{vonkorff2014}, where within a few hundred fs
magnetization reduction has a spatial distribution.  In our picture,
magnetic and electric fields of the laser pulse jointly drive the
electron down along the axial direction as a wavepacket (see
Fig. \ref{fig1}).

Our approach is in sharp contrast to the prior studies, where 
the axial motion is empirically included \cite{battiato2010} or
simulated by chemical potential
\cite{choi2014,fognini2017,kimling2017}.  Indeed, Ashok and coworkers
\cite{ashok2022} started with a space- and time-dependent transient
chemical potential and computed the internal electric field from the
chemical potential and then current using the Ohm's law.  Hurst \et
\cite{hurst2018} recognized the importance of the magnetic field and
adopted the spin-Vlasov equation, instead of the Boltzmann
equation. To simulate the external electric field effect, they shifted
the distribution by a constant velocity $\Delta v$.  Choi \et
\cite{choi2014,choi2014b} gave the spin accumulation through the spin
diffusion equation as \be \frac{\partial \mu_s}{\partial t}=D
\frac{\partial^2 \mu_s}{\partial z^2}-\frac{\mu_s}{\tau_s}, \ee where
$\mu_s=\mu_\uparrow -\mu_\downarrow$ is the spin chemical potential
and D is the spin diffusion constant, and $\tau_s$ is the spin
relaxation time. This approach follows the electric-field induced
transport, where a voltage bias is applied across a device and the chemical
potential difference maintained between the drain and source drives
the charge carrier across the device \cite{datta}. So the
charge carriers move  along the electric field. In a shorter
channel, the transport can be ballistic, while in a longer channel, it
is diffusive.

Our finding has several important implications.  First, although the
magnetic field is much weaker than the electric field, if we ignore
it, the electron only moves along the $x$ axis. Therefore, it is
absolutely necessary to include the magnetic field, because it is the
magnetic field that changes the direction of the electron motion
\cite{hagenbuch1977}.  This is also consistent with the Poynting
vector, thus, the momentum, which is in the direction of light
propagation. Although ignoring the magnetic field greatly eases
numerical calculation, it cannot describe the laser-induced electron
transport. Second, because ${\bf B}=\nabla \times {\bf A}$, where
${\bf A}$ is the vector potential, not to be confused with the laser
field amplitude $A_0$.  A nonzero ${\bf B}$ means that ${\bf A}$ must
contain $\br$ as its variable. The regular dipole approximation, ${\bf
  A}\rt \rightarrow {\bf A}(t)$, is not appropriate, if one wishes to
describe the electron transport along the axial direction, but nearly
all prior theoretical studies employ this approximation
\cite{battiato2010,jap11,krieger2015,chen2019w,prl20}. For instance,
the first model simulation \cite{prl00} and even more recent
first-principles investigations \cite{jap11,krieger2015,prl20} used a
vector potential ${\bf A}$ that is independent of space. Such an
approach, called the dipole approximation, is reasonable if the
wavelength of a laser pulse is much larger than the area illuminated
and the focus is on the electronic excitation, rather than transport.
Without a spatial dependence, the electron only moves along the ${\bf
  E}$ field direction \cite{jpcm18} which is transversal to the light
propagation direction and the oscillating current is exactly in the
same direction as the polarization \cite{foldi2013}.


In conclusion, we have shown that the electron transport under an
ultrafast laser pulse is most likely due to the joint effect of the
electric and magnetic fields. Each field alone cannot introduce the
transport along the axial direction. The electric field provides a
strong transversal velocity, while the magnetic field steers the
electron moving along the light propagation direction.  With the
experimental accessible laser field parameters, we demonstrate that
the electron can move up to 20 to 262 $\rm\AA$, which now agree with
the prior experiments \cite{brorson1987,melnikov2011}. Our finding
points out a serious problem with the existing theories where the
vector potential has no spatial dependence but is used to describe the
transport. Our theory goes beyond the classical Boltzmann equation and
chemical potential change, and puts the transport physics on a solid
foundation.  This potentially opens the door to more advanced
first-principles theory, which has a significant impact on future
technology in charge and spin transports
\cite{brorson1987,melnikov2011,stanciu2007,mplb18,aip20,liu2020,liu2023,kang2023}.

\acknowledgments

This work was partially supported by the U.S. Department of Energy
under Contract No. DE-FG02-06ER46304. Part of the work was done on
Indiana State University's high performance Quantum and Obsidian
clusters.  The research used resources of the National Energy Research
Scientific Computing Center, which is supported by the Office of
Science of the U.S. Department of Energy under Contract
No. DE-AC02-05CH11231.

$^*$ guo-ping.zhang@outlook.com.
 https://orcid.org/0000-0002-1792-2701

The data that support the findings of this study are available from
the corresponding author upon reasonable request.

\end{document}